\documentclass[twocolumn,showpacs,preprintnumbers,amsmath,amssymb,superscriptaddress,10pt,aps,prb]{revtex4-1}
\usepackage{epsfig,amsopn}
\usepackage{graphicx}
\usepackage{epstopdf}
\usepackage{amsmath,amssymb}
\usepackage{amsthm}
\usepackage{enumerate}
\usepackage{xcolor}

\begin{document}
\title{Spin Response and Collective Modes in Simple Metal Dichalcogenides}

\author{Dibya Kanti Mukherjee}
\affiliation{Harish-Chandra Research Institute, HBNI, Chhatnag Road, Jhunsi, Allahabad 211 019, India}
\author{Arijit Kundu}
\affiliation{Department of Physics, Indian Institute of Technology Kanpur, Kanpur 208016, India}
\author{H.A. Fertig}
\affiliation{Department of Physics, Indiana University, Bloomington, IN 47405}
	
\begin{abstract}
Transition metal dichalcogenide (TMD) monolayers are interesting materials in part
because of their strong spin-orbit coupling. This leads to intrinsic spin-splitting of opposite
signs in opposite valleys, so the valleys are intrinsically spin-polarized when
hole-doped.   We study spin response in a simple model of these materials,
with an eye to identifying sharp collective modes (i.e, spin-waves) that are
more commonly characteristic
of ferromagnets.  We demonstrate that such modes exist for arbitrarily weak repulsive interactions,
even when they are too weak to induce spontaneous ferromagnetism.
The behavior of the spin response is explored for a range of hole dopings and interaction strengths.
\end{abstract}
	
\maketitle
	
\section{Introduction}
Two-dimensional materials based on honeycomb lattices have become a subject of intense
investigation in the past few years, due to their interesting band structure and associated
topological properties.  The low-energy dynamics of such systems are typically dominated
by states near the $K$ and $K'$ points in the Brillouin zone.  The paradigm of this
is realized in graphene, a pure carbon honeycomb lattice, which hosts a gapless spectrum with
Dirac points at these locations \cite{Castro_Neto_RMP} due to a combination
of inversion and time-reversal symmetry, as well as the very weak spin-orbit coupling (SOC)
typical of light elements.  More recently, transition metal dichalcogenide (TMD) monolayers,
where a transition metal $M$ (e.g., Mo or W) resides on one sublattice and a dimer of
chalcogen $X$ atoms (e.g., S, Se) on the other,
have emerged as important materials in this class \cite{Mak_2010,Splendiana_2010}.
These system are gapped at the $K$ and $K'=-K$ points,
and the strong SOC associated with $M$ atoms leads to very interesting
spin-valley coupling near these points \cite{Xiao_2012,Liu_2013}.
In particular, one finds spin up and down
components of the valence band well-separated in energy, with their ordering interchanged
for the two valleys. This allows for an effective valley polarization to be induced when the
system spin polarizes via pumping with circularly polarized light \cite{Cao_2012,Zeng_2012,Mak_2012}.
The coupling of spin and valley in this way has been dramatically demonstrated via the
observation of a valley Hall effect in this circumstance \cite{Mak_2014}.

The locking of spin and valley degrees of freedom in TMD monolayers is a unique feature of these materials.
When hole-doped, it leads to a non-zero expectation value of $\sigma_z\tau_z$,
where $\sigma_z$ a Pauli matrix for spin, and $\tau_z$ the analogous operator for the valley index.
This occurs without any interaction present in the Hamiltonian, yet is reminiscent of ferromagnetic
ordering, albeit without time-reversal symmetry-breaking since this reverses both spin
and valley.  Recently, it has been argued that for strong enough interactions, TMD systems
develop a spontaneous imbalance of spin/valley populations \cite{Scrace_2015,Braz_2017},
which leads to actual ferromagnetic spin order in the groundstate.  It thus becomes
interesting to consider how one might probe and distinguish these orderings.  One possible
strategy is to investigate the spin response of the system, both to search for sharp
collective modes that are a hallmark of ferromagnets, and to understand broader features of
the response that demonstrate the ordering present in these materials.  This is
the subject of our study.

\begin{center}
	\begin{figure}[t]
		\includegraphics[width=0.44\textwidth]{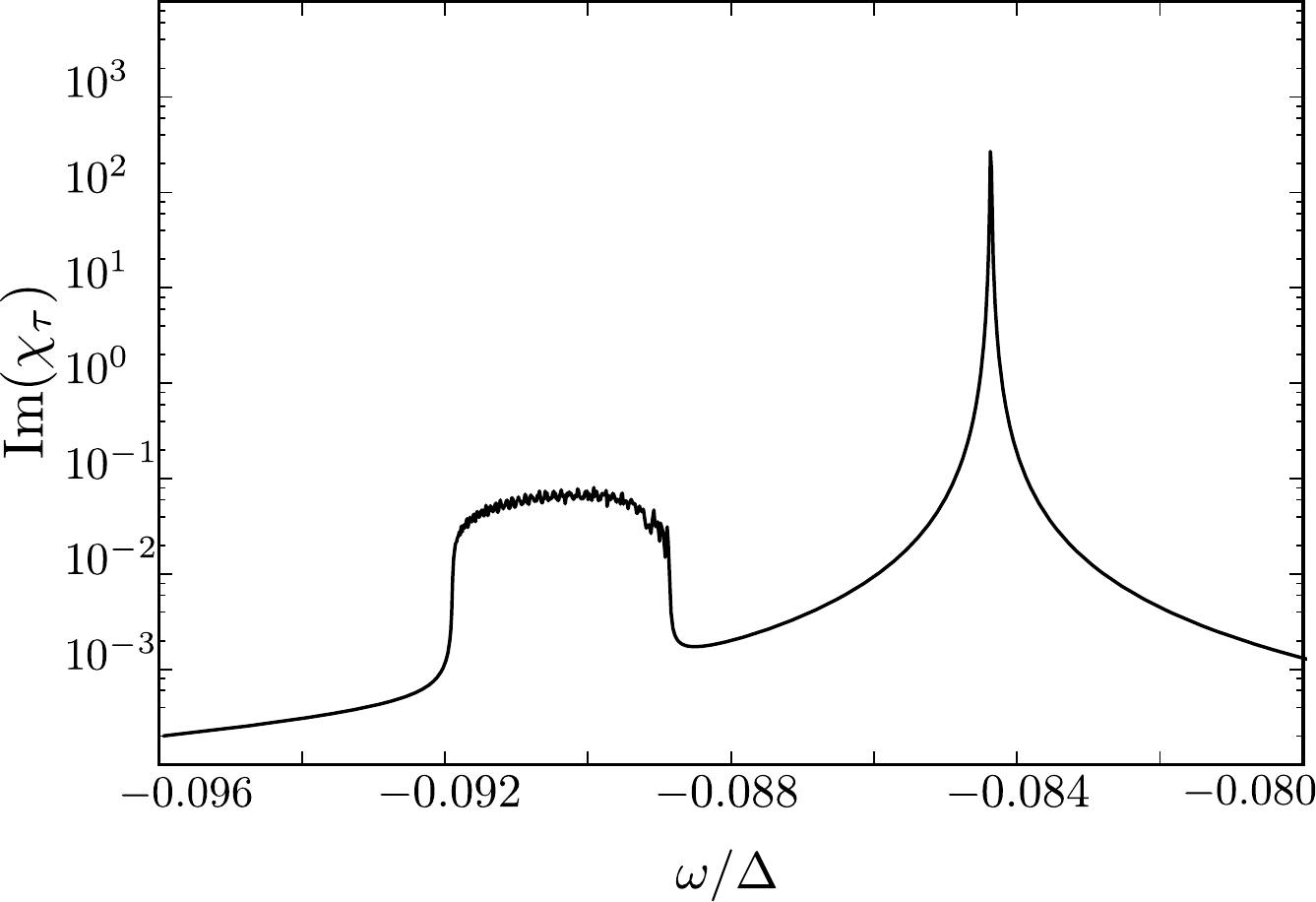}
		\caption{Absorptive part of spin response function
Im $\chi_\tau({\bf q},\omega)$
for $\textbf{q}=0$, chemical potential $\mu_0=-0.49\Delta$ and $U_0=0.2$eV with $\tau=+1$.  Model parameters for band structure in Table I.
A sharp collective mode near $\omega \approx -0.0845\Delta$ is prominent above a particle-hole continuum
in the interval $-0.092 \lesssim \omega/\Delta \lesssim -0.087$, where $\Delta$ = 1.66 eV.
}
\label{fig:phys_response}
	\end{figure}
\end{center}

We focus on the basic qualitative physics of this system by employing a simple two-band model
for $MX_2$ compounds \cite{Xiao_2012} with a short-range repulsive interaction, and compute the spin response
using the time-dependent Hartree-Fock approximation (TDHFA) \cite{Giuliani_book}.  For concreteness
quantitative results are computed using parameters appropriate for MoS$_2$,
and we examine results for several representative hole-dopings
and interaction strengths.  A typical result is illustrated in Fig. \ref{fig:phys_response} for a
system with low hole doping, such that only a single spin species of the valence band is partially
unoccupied in each of the valleys.

For small wavevectors
$q$, a sharp collective mode is visible below a continuum of particle-hole spin-flip excitations
which are present even in the absence of interactions (although the frequency interval where
they reside is renormalized by them).  An interesting feature of the
collective mode is that, for low hole doping, it is present for arbitrarily weak interaction strength,
even if
the system is not spin-ferromagnetic.  Its presence may be understood as arising from the
effective $\sigma_z\tau_z$ polarization that is induced when the system is hole-doped.
Interestingly, this is a direct analog of ``Silin-Leggett'' modes \cite{silin_1958,leggett_1970}
that appear when fermions become spin-polarized by a external magnetic field. In that system, the non-interacting Hamiltonian induces a spin polarization in the groundstate which is not present spontaneously.  Nevertheless, the combination of different Fermi surfaces for different spins, together with exchange interactions which energetically favor ferromagnetism locally, leads to sharp, collective excited states of low energy.  These modes have been detected in spin-polarized $^3$He \cite{Tastevin_1985}.

In the TMD system,
an analogous sharp response appears when the system absorbs angular momentum, typically
from a photon, and is dominated by excitations around one of the two valleys.  The spin response
from the other valley is negligible around these frequencies, but can be seen at negative
frequencies, which is equivalent to absorption of photons with the opposite helicity.
This effect is well-known in the context of undoped TMD systems \cite{Zeng_2012, Mak_2012,Cao_2012}
where the particle-hole excitations involve electrons excited from the valence to the conduction
band.  In the present situation one finds this behavior from excitations within the valence band,
from occupied spin states to unoccupied ones available due to the doping, of opposing spin.
The resulting sharp modes are much lower in energy than comparable exciton modes of an
undoped system \cite{Ross_2013,Ugeda_2014,Wu_2015,Trushin_2016}.

True ferromagnetism in this system has been argued to arise when interactions are sufficiently strong that unequal populations of the two valleys becomes energetically
favorable \cite{Scrace_2015,Braz_2017}, and for a hole-doped, short-range interaction model, it occurs as a
first-order transition at a critical interaction strength $U_c \cite{Braz_2017}$.  Within our model this results in an effective shift of the bands relative to one another, so that a system sufficiently clean and cold to allow observation of resonances associated with collective spin modes
would present them at different frequencies for different helicities.

\begin{center}
	\begin{figure}[t]
		\includegraphics[width=0.45\textwidth]{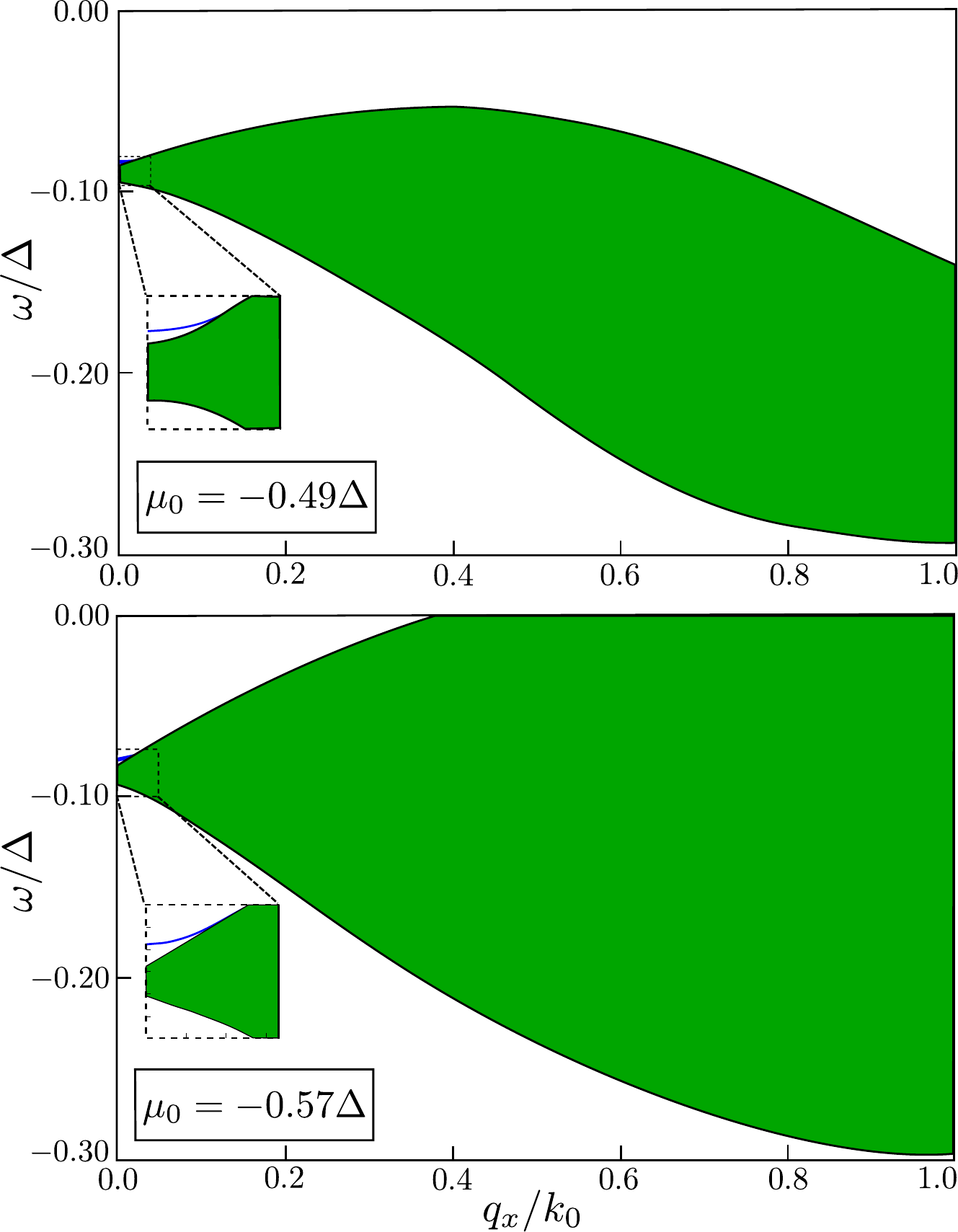}
		\caption{The top panel, for $\mu_0=-0.49\Delta$, in which there is only a single
Fermi surface in the valley (demonstrated in Fig.~\ref{fig:band}), has a continuum of particle-hole
excitations (shown in green) below some minimum frequency.
The lower panel has $\mu_0=-0.57\Delta$ for which there are two Fermi surfaces in the valley, giving rise to the continuum modes with vanishingly small energies for $q_x>0.4k_0$
with $k_0=\Delta/2ta$. For both panels, $U_0=0.2$eV and $\tau=+1$. Other parameters are listed in Table I.
Blue lines illustrate the collective spin wave mode dispersion.}
\label{fig:loww}
	\end{figure}
\end{center}

At higher dopings the valence bands will support two Fermi
surfaces in each valley, indicating that they contain holes of both spins.
Because of the opening of the second Fermi surface the system now supports
gapless spin-flip excitations, albeit at finite wavevector.
Regions in
frequency and wavevector where these exist are illustrated in Fig.
\ref{fig:loww}, along with the spin wave dispersion for these parameters.
Observation of such a continuum of gapless modes would allow a direct
demonstration of the spin-split Fermi surfaces in this system.  In practice,
because these modes appear above wavevectors of order
$q \lesssim 1/a$ with $a$ the lattice constant, their
presence may be difficult to observe by direct electromagnetic absorption
because of momentum conservation.  In real systems,
disorder relaxes this constraint and may make their detection feasible \cite{Pinczuk_1997}.

Our analysis also shows that the system in principle supports a {\it second}
collective spin wave mode, one associated
with inter-orbital spin flips.  This mode exists extremely close to the edge of the
continuum of particle-hole spin excitations and in practice might be difficult to discern in the spin-response
function.  Its presence would presumably be more easily detected in response functions
that combine inter-orbital excitations with spin flips.

This article is organized as follows.  In Section II we describe both the single particle
Hamiltonian and the interaction model we adopt for this system.  Section III describes
a static Hartree-Fock analysis of the system, demonstrating that the effective single-particle
Hamiltonian is rather similar to the non-interacting one, with renormalized parameters.
In Section IV we carry out a time-dependent Hartree-Fock analysis of the spin response
function, and show how one can identify poles that signal allowed spin-flip excitations of
the system.  In Section V we carry out an analytic analysis of the equations generated in
the previous section, appropriate for low hole doping.  Section VI provides results one
finds from numerical solutions for the spin response functions.  We conclude with a summary
in Section VII.

\section{Model of the system}	
Our starting point is a simple two-band Hamiltonian for the monolayer MX$_2$, such as MoS$_2$, developed through several numerical, symmetry-based analyses~\cite{Xiao_2012} which capture the electronic properties near the $K,-K$ valleys.  In the absence of interactions this has the form
\begin{align}
H_0^\tau(\textbf{k}) =\left[
\begin{array}{cc}
\Delta/2       & at(\tau k_x-ik_y) \\
at(\tau k_x+ik_y)       &  -\Delta/2 +s\tau\lambda\\
\end{array}\right],\label{eq:ham}
\end{align}
which is written in the basis $|\psi^{\tau}_c\rangle = |d_{z^2}\rangle$ and $|\psi^{\tau}_v\rangle = \frac{1}{\sqrt{2}}(|d_{x^2-y^2}\rangle +i\tau|d_{xy}\rangle)$, where $\tau=\pm$ is the valley index, $t$ is the hopping matrix element and $d_{z^2}$, $d_{x^2-y^2}$, $d_{xy}$ are orbitals of the $M$ atoms.
(Here and throughout this paper we take $\hbar = 1$.)
Spin is a good quantum number, denoted by $s=1$ for $\uparrow$ and $s=-1$ for $\downarrow$. The strength of spin-orbit coupling is encoded
in the parameter
$\lambda$.  In the ground state of this Hamiltonian, states up to the chemical potential $\mu_0$,
which is tunable in principle via gating,
are filled. Estimates
\cite{Xiao_2012} for the parameters relevant to MoS$_2$ are listed in Table I.

\begin{figure}[t]
  \includegraphics[width=0.48\textwidth]{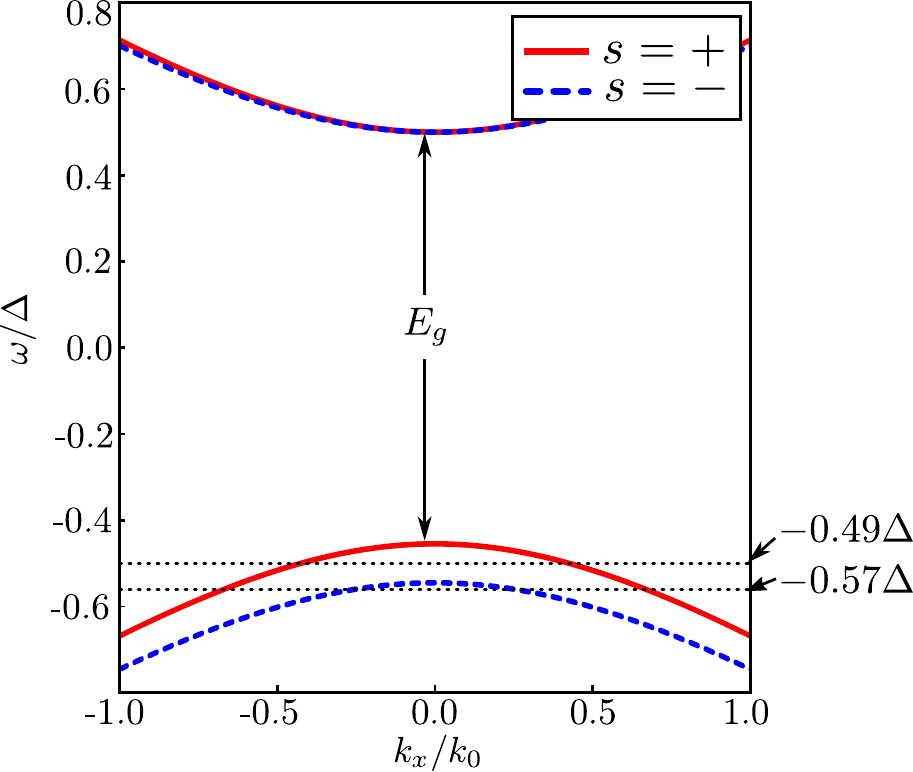}
  \caption{The band dispersion of Hamiltonian (\ref{eq:ham}) showing a direct band gap $E_g$ between the valence and the conduction band and the separation of spin polarized bands in the conduction band. Position for two of $\mu_0$ are marked on the right margin. $k_0=\Delta/2ta$ is the scale of momentum. The parameters used are listed in Table 1 and $\tau=+1$.}\label{fig:band}
\end{figure}

\begin{center}
	\begin{table}[h]
	\begin{tabular}{ | c | c | c | c |}
		\hline
		$a$ & $t$ & $\Delta$ &  $\lambda$ \\ \hline
		 3.190 \r{A}  & 1.059 eV & 1.66 eV & 0.075 eV \\
		\hline
	\end{tabular}
\caption{Values of various parameters for MoS$_2$ from Ref. \onlinecite{Xiao_2012}.}
\end{table}
\end{center}
The energy eigenstates of the full Hamiltonian with momentum $\textbf{k}$ and spin $s$ will be denoted by $\phi_{l,s}(\textbf{k})$, with $l=\{\tau,\alpha\}$ ($\alpha = \pm$ for conduction/valence bands),
and have the form
\begin{align}\label{freephi}
\phi_{l,s}(\textbf{k}) = \frac{1}{\sqrt{2}}\left(\begin{array}{c} \tau e^{-i\tau\phi}\sqrt{1+\frac{\alpha m_{s\tau}}{\sqrt{m_{s\tau}^2+a^2t^2k^2}}} \\ \alpha\sqrt{1-\frac{\alpha m_{s\tau}}{\sqrt{m_{s\tau}^2+a^2t^2k^2}}} \end{array}\right),
\end{align}
with corresponding eigenvalues
\begin{align}\label{eq:freeen}
&\epsilon_{l,s}^{\alpha}(\textbf{k}) =\frac{\tau s\lambda}{2} + \alpha \sqrt{m_{s\tau}^2+(atk)^2},
\end{align}
where $m_{s\tau}=\frac{\Delta-\tau s\lambda}{2}$ and $k=\sqrt{k_x^2+k_y^2}$. The bands near the $K$
($\tau=1$) valley, shown in Fig.~\ref{fig:band}, illustrate the distinct spin structure of the system.
The valence and conduction band are separated by a relatively large
gap $E_g=(\Delta -\lambda)$ at $k=0$, whereas the two spin valence bands are further separated by a
smaller gap of magnitude $E_{\lambda}=2\lambda$. This gap between the spin-split valence bands remains almost constant for a range of $k$ until $akt\gg \Delta$.  Note that the two conduction bands
of the model are nearly degenerate.  The $K$ and $-K$ valleys of the system are related by time-reversal,
so that the spins of the two bands are reversed in going from one to the other.

To write down an effective interaction, it is convenient to define field operators of
spin $s$ projected into the set of states defined in our model,
\begin{align}
\Psi_s(\textbf{r}) = \frac{1}{\sqrt{L_xL_y}}\sum_{\textbf{k},l}
e^{i(\textbf{k}+\textbf{K}_{\tau_l})\cdot \textbf{r}}\phi_{l,s}(\textbf{k})c_{l,s}(\textbf{k}),
\end{align}
where $c_{l,s}(\textbf{k})$ is the annihilation operator for the $l,s$ state at momentum $\mathbf{k}$
relative to the valley minima/maxima at $\textbf{K}_{\tau_l} = \tau_l \textbf{K}$, with the sign determined
by the $\tau$ index implicit in $l$, and $L_xL_y$ is the area of the system.
A repulsive interaction among the band-electrons can then be represented in the form
\begin{widetext}
\begin{align}
H_{\text{int}} = {1 \over 2} \sum_{s,s'} \int d^2\textbf{r} d^2\textbf{r}' V(\textbf{r}-\textbf{r}'):\Psi^\dagger_{s}(\textbf{r})\Psi_{s}(\textbf{r})
\Psi^\dagger_{s'}(\textbf{r}')\Psi_{s'}(\textbf{r}'):,
\end{align}
with $V$ represents a finite-range repulsive interaction.
Physically this arises from Coulomb interactions among the band electrons;
the finite range can be provided by a screening gate or by carriers in the
layer itself (although we will not treat the screening dynamically in what follows).
We assume the screening length is large on the scale of the lattice constant
so that inter-valley contributions to the density $\Psi^\dagger_{s}(\textbf{r})\Psi_{s}(\textbf{r})$
oscillate rapidly, and can be ignored when integrated over ${\bf r}$.  This leads to the
replacement
\begin{align}
H_{\text{int}} \rightarrow {1 \over 2} \sum_{s,s'} \sum_{\tau,\tau'}
\int d^2\textbf{r} d^2\textbf{r}' V(\textbf{r}-\textbf{r}')
:\Psi^\dagger_{s\tau}(\textbf{r})\Psi_{s\tau}(\textbf{r})
\Psi^\dagger_{s'\tau'}(\textbf{r}')\Psi_{s'\tau'}(\textbf{r}'):,
\end{align}
with
\begin{align}
\Psi_{s\tau}(\textbf{r}) = \frac{1}{\sqrt{L_xL_y}}\sum_{\textbf{k},l}
e^{i\textbf{k}\cdot \textbf{r}}\phi_{l,s}(\textbf{k})c_{l,s}(\textbf{k})\delta_{\tau,\tau_l},
\end{align}
where $\tau_l$ is the valley content of the composite $l$ index.  At this point we can
make the approximation
$V(\mathbf{r}-\mathbf{r}') = 2U_0 \delta^2(\mathbf{r}-\mathbf{r}')$, and
arrive at an interaction form
\begin{align}
H_{\text{int}} = &U \sum_{\{l_i \textbf{k}_i \textbf{q}\}}\sum_{s,s'}
\phi^\dagger_{l_1 s}(\textbf{k}_1)\phi^\dagger_{l_2 s'}(\textbf{k}_2)
\phi_{l_3 s'}(\textbf{k}_2+\textbf{q}')\phi_{l_4 s}(\textbf{k}_1-\textbf{q}')
\delta_{\tau_{l_1},\tau_{l_4}} \delta_{\tau_{l_2},\tau_{l_3}}
c^\dagger_{l_1 s}(\textbf{k}_1) c^\dagger_{l_2 s'}(\textbf{k}_2)
c_{l_3 s'}(\textbf{k}_2+\textbf{q}')c_{l_4 s}(\textbf{k}_1-\textbf{q}'),
\label{Hint}
\end{align}
where $U=\frac{U_0}{L_xL_y}$.  This is the interaction Hamiltonian that we use
in the Hartree-Fock analyses that follow.
\vskip 4mm


\end{widetext}

\begin{center}
	\begin{figure}[t]
		\includegraphics[width=0.45\textwidth]{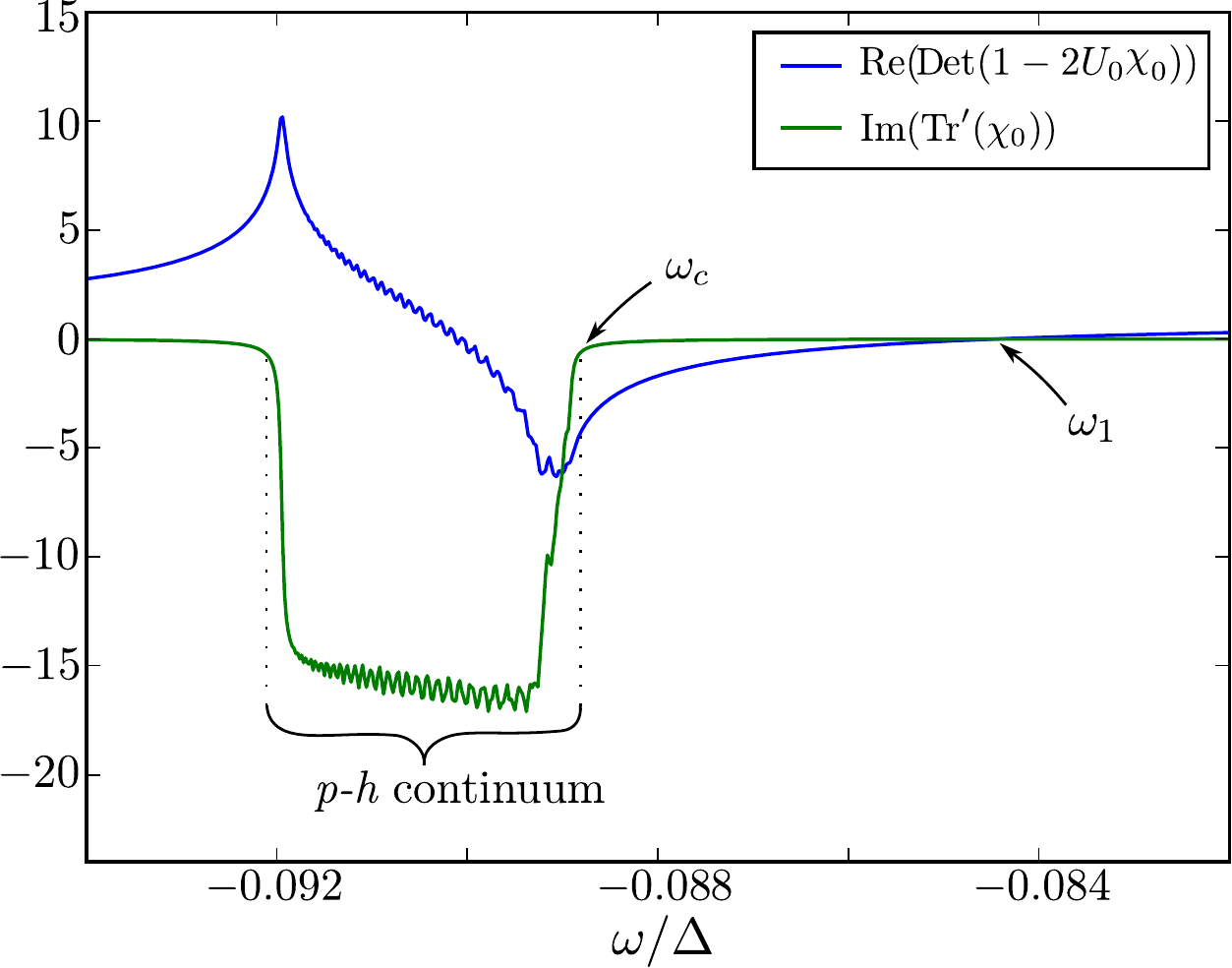}
		\caption{Plot of a typical $\chi(\mathbf{q},\omega)$, Eq.~(\ref{eq:finalchi}), showing the particle-hole excitations of the spin-split valence bands below an energy $\omega_c$. At $\omega_1$, there is a single collective mode visible for which the real part of the denominator of Eq.~(\ref{eq:finalchi}) is zero. Here we have used $\mathbf{q} = \mathbf{0}$, $\mu_0=-0.49\Delta$, $\tau=+1$ and $U_0=0.2$eV.}\label{fig:chi}
	\end{figure}
\end{center}
\section{Hartree-Fock Approximation}
In order to carry out an analysis of the spin response in this system within
the time-dependent Hartree-Fock approximation, it is first necessary to find
the density matrix of the system within the static Hartree-Fock (HF) approximation.
This has the form
\begin{equation}
	\langle c^\dagger_{ls}(\textbf{k})c_{l's'}(\textbf{k}') \rangle = n_{ls}(\textbf{k})\delta_{ll'}\delta_{ss'}\delta_{\textbf{k},\textbf{k}'}.
\end{equation}
Note in writing this, we have assumed that neither interband nor intervalley coherence
have formed in the system spontaneously.  Performing a HF decomposition on Eq.~(\ref{Hint}) gives a potential for an effective single-body Hamiltonian,
\begin{align}
H^{\text{HF}}_{\text{int}} =& -2U\sum_{ll',ss',\textbf{k}}\delta_{ss'}\sum_{a,b=A/B}c^{\dagger}_{ls}\phi^{a*}_{ls}({\textbf{k}})\times\nonumber\\
&\times\left(\sum_{l''}\phi_{ls}^a(\textbf{k})n_{l''s}(\textbf{k})\phi_{l''s}^{b*}(\textbf{k})\right)\phi^{b}_{l's}({\textbf{k}})c_{l's}.\label{eq:hfint}
\end{align}
where, for notational simplicity, we have used the $a,b$ indices to denote the orbital degree of freedom ($A \equiv   |d_{z^2}\rangle$ and $B \equiv \frac{1}{\sqrt{2}}(|d_{x^2-y^2}\rangle +i\tau|d_{xy}\rangle) $). The full HF Hamiltonian
for electrons with wavevector ${\bf k}$ then becomes
\begin{equation}
H^{0,\text{HF}}_{ls,l's}(\textbf{k})=H^{0}_{ls,l's}(\textbf{k}) - 2U\sum_{ab}\phi^{a*}_{ls}({\textbf{k}})n^{ab}_s\phi^{b}_{l's}({\textbf{k}}),
\end{equation}
with $n^{ab}_{s \tau_l} = \sum_{\textbf{k}l}\phi_{ls}^a(\textbf{k})n_{ls}(\textbf{k})\phi_{ls}^{b*}(\textbf{k})$.
The quantities $n_{ls}$ need to be determined self-consistently.
Note in writing $H^{0,\text{HF}}_{ls,l's}(\textbf{k})$, we have dropped
a term proportional to the total fermion number which is a constant. In the orbital basis ($l,l'$) one may write
\begin{equation}
	H^{0,\text{HF}}(\textbf{k}) =\left[
	\begin{array}{cc}
		\tilde{m}_{s\tau}       & at\tau k e^{-i\tau\phi} \\
		at\tau k e^{i\tau\phi}       &  -\tilde{m}_{s\tau}\\
	\end{array}\right]+\tau s \lambda/2 - U(n^{AA}_{s\tau} + n^{BB}_{s\tau}),\label{eq:hf0}
\end{equation}
with renormalized mass
$\tilde{m}_{s \tau} = \frac{\Delta-\tau s\lambda}{2} - U(n^{AA}_{s \tau} - n^{BB}_{s \tau})$. For a fixed density (obtained by fixing $\mu_0$), the value of $\tilde{m}_{\tau s}$ is found
numerically using the requirement that the values $n_{ls}(k)$ used to generate Eq.~(\ref{eq:hf0})
yield wavefunctions that produce the very same values -- i.e., the density matrix
used to generate the HF Hamiltonian is the same as what one finds from its
eigenvectors and eigenvalues.  In the present case, the wavefunctions have
a functional form that is the same as that of
the free wavefunctions, Eq.~(\ref{freephi}),
with modified parameters:
\begin{align}\label{hfphi}
\phi_{l,s}(\textbf{k}) = \frac{1}{\sqrt{2}}\left(\begin{array}{c} \tau e^{-i\tau\phi}\sqrt{1+\frac{\alpha \tilde{m}_{s\tau}}{\sqrt{\tilde{m}_{s\tau}^2+a^2t^2k^2}}} \\ \alpha\sqrt{1-\frac{\alpha \tilde{m}_{s\tau}}{\sqrt{\tilde{m}_{s\tau}^2+a^2t^2k^2}}} \end{array}\right).
\end{align}
The energy eigenvalues then become
\begin{align}
&\tilde{\epsilon}_{l,s}(\textbf{k}) =\frac{\tau s\lambda}{2} + \alpha \sqrt{\tilde{m}_{s\tau}^2+(atk)^2}  - U(n^{AA}_{\tau s} + n^{BB}_{\tau s}),\label{eq:HFE}
\end{align}
which is similar but not identical to the non-interacting energy eigenvalues, Eq.~(\ref{eq:freeen}). Here, in analogy with the previous section, the index $l=\{\tau,\alpha\}$
implicitly contains the valley index $\tau$ as well as the
conduction/valence band index $\alpha=\pm1$. In the remainder of this paper,
we will use these as the basis states for our analysis.

\section{Time dependent Hartree-Fock Approximation}
Our focus in this study is the spin-spin response function
\begin{align}
\chi_{\tau}(\textbf{r}-\textbf{r}',t) =  -i\Theta(t)\langle[\rho^{+-}_{\tau} (\textbf{r},t), \rho^{-+}_{\tau} (\textbf{r}',0)] \rangle,\label{eq:corr}
\end{align}
with $\rho^{\sigma\sigma'}_{\tau}(\mathbf{r},t) = \Psi^{\text{HF}\dagger}_{\sigma\tau}(\textbf{r},t)\Psi^{\text{HF}}_{\sigma'\tau}(\textbf{r},t)$, with field operators
\begin{align}
\Psi_{s\tau}^{\text{HF}}(\textbf{r}) = \frac{1}{\sqrt{L_xL_y}}\sum_{\textbf{k},l}
e^{i\textbf{k}\cdot \textbf{r}}\phi_{l,s}(\textbf{k})c_{l,s}(\textbf{k})\delta_{\tau,\tau_l}.
\end{align}
The single particle states appearing in this expression are the HF wavefunctions, Eq.~(\ref{hfphi}). We do not consider intervalley particle-hole operators as this would involve large momentum imparted to the system. Assuming translational invariance, in momentum space the response function has the form
\begin{widetext}
\begin{align}
\chi_{\tau}(\textbf{q},t)
=-&\frac{i\Theta(t)}{L_xL_y}\sum_{\{\textbf{k}_i,\textbf{q}_i,l_i\}}f_{l_1l_2,\uparrow\downarrow}(\textbf{k}_1+\textbf{q},\textbf{k}_1)f_{l_3l_4,\downarrow\uparrow}(\textbf{k}_2-\textbf{q},\textbf{k}_2)  \langle[e^{iHt}c^\dagger_{l_1\uparrow}(\textbf{k}_1+\textbf{q})c_{l_2\downarrow}(\textbf{k}_1)e^{-iHt},c^\dagger_{l_3\downarrow}(\textbf{k}_2-\textbf{q})c_{l_4\uparrow}(\textbf{k}_2)]\rangle \nonumber \\
\equiv & \frac{1}{L_xL_y}\sum_{\{\textbf{k}_i,\textbf{q}_i,l_i\}}f_{l_1l_2,\uparrow\downarrow}(\textbf{k}_1+\textbf{q},\textbf{k}_1)f_{l_3l_4,\downarrow\uparrow}(\textbf{k}_2-\textbf{q},\textbf{k}_2)
\tilde{\chi}_{l_1l_2l_3l_4}(\textbf{k}_1,\textbf{k}_2,\textbf{q},t), \label{eq:physresp}
\end{align}
with
\begin{align}
\tilde{\chi}_{l_1l_2l_3l_4}(\textbf{k}_1,\textbf{k}_2,\textbf{q},t) =-i\Theta(t) \langle[e^{iHt}c^\dagger_{l_1\uparrow}(\textbf{k}_1+\textbf{q})c_{l_2\downarrow}(\textbf{k}_1)e^{-iHt},c^\dagger_{l_3\downarrow}(\textbf{k}_2-\textbf{q})c_{l_4\uparrow}(\textbf{k}_2)]\rangle. \label{eq:chitilde}
\end{align}
It is implicit that the $\tau_{l}$ content of each $l$ index on the right hand side
of this equation is a single value of $\tau$, and the Hamiltonian
appearing in the $e^{\pm iHt}$ factors
is $H = H_0+ H_{\text{int}}$, using Eqs.~(\ref{eq:ham}) and (\ref{Hint}). The
weights $f_{l_il_j,\sigma\sigma'}(\mathbf{k}_1,\mathbf{k}_2) \equiv
\phi_{l_i\sigma}^{\dagger}(\mathbf{k}_1)\phi_{l_j\sigma'}(\mathbf{k}_2)$ are
wavefunction overlap factors, and the indices $l_i$ have allowed values
$\tau_l=\pm1$ and $\alpha_l = \pm1$.
To obtain an explicit expression for $\tilde{\chi}$,
we take a time derivative of its definition implicit in Eq.~(\ref{eq:physresp}), which
generates expectation values involving 2, 4, and 6 fermion operators.  We approximate
the last of these using a HF decomposition \cite{Giuliani_book}, leading to a closed
expression for the response function that involves elements of the static density
matrix described in the last subsection.  This is the form in which we carry out
the time-dependent Hartree-Fock approximation.  The resulting equation may be
expressed as
\begin{align}
    i\partial_t\tilde{\chi}_{l_1l_2l_3l_4}(\textbf{k}_1,\textbf{k}_2,\textbf{q},t) =& \{n_{l_1\uparrow}(\textbf{k}_1+\textbf{q})-n_{l_2\downarrow}(\textbf{k}_1)\}\delta_{l_1l_4}\delta_{l_2l_3}\delta_{\textbf{k}_1,\textbf{k}_2-\textbf{q}}  - \Big[\tilde{\epsilon}_{l_1,\uparrow}(\textbf{k}_1+\textbf{q}) - \tilde{\epsilon}_{l_2,\downarrow}(\textbf{k}_1)\Big] \tilde{\chi}_{l_1l_2l_3l_4}(\textbf{k}_1,\textbf{k}_2,\textbf{q},t)\nonumber  \\
& + 2U\sum_{ab}\Big[\phi^a_{l_1\uparrow}(\textbf{k}_1+\textbf{q})\Big(n_{l_2\downarrow}(\textbf{k}_1)-n_{l_1\uparrow}(\textbf{k}_1+\textbf{q})\Big)\phi^{b*}_{l_2\downarrow}(\textbf{k}_1)\Big]\tilde{\chi}^{ab}_{\uparrow\downarrow l_3l_4}(\textbf{k}_1,\textbf{k}_2,\textbf{q},t),
\label{deriv}
\end{align}
where
$$ \tilde{\chi}^{ab}_{s_1s_2 l_3l_4}(\textbf{k}_2, \textbf{q}, t) \equiv \sum_{l_1l_2\textbf{k}_1}\phi^{a*}_{l_1s_1}(\textbf{k}_1+\textbf{q})\phi^{b}_{l_2s_2}(\textbf{k}_1)\tilde{\chi}_{l_1l_2l_3l_4}(\textbf{k}_1,\textbf{k}_2,\textbf{q},t) $$
defines $\tilde{\chi}^{ab}_{\uparrow \downarrow l_3 l_4}$
and $\phi^{a}_{l,s}$ is the amplitude for the $a$th orbital (see Eq.~(\ref{hfphi})). Some details leading up to Eq.~(\ref{deriv}) are provided in Appendix A.
Fourier transforming Eq.~(\ref{deriv}) with respect to time,
with further work it may be cast in the form
\begin{align}
-\chi_0^{cd,c'd'}(\textbf{q},\omega) = \chi^{cd,c'd'}(\textbf{q},\omega) - 2U_0\sum_{ab}\chi_0^{cd,ab}(\textbf{q},\omega)\chi^{ab,c'd'}(\textbf{q},\omega).\label{eq:chi}
\end{align}
Here $U_0 = L_xL_y U$, $\chi^{cd,c'd'}(\textbf{q},\omega) \equiv \frac{1}{L_xL_y}\sum_{l_3,l_4,\textbf{k}}\tilde{\chi}^{cd}_{\uparrow\downarrow l_3l_4}(\textbf{k},\textbf{q},\omega)
\phi^{c'}_{l_4\uparrow}(\textbf{k})\phi^{d'*}_{l_3\downarrow}(\textbf{k}-\textbf{q})$, and
\begin{equation}
\chi_0^{ab,cd}(\textbf{q}, \omega) = - \frac{1}{L_xL_y}\sum_{l_3,l_4,\textbf{k}_2} \frac{n_{l_4\uparrow}(\textbf{k}_2)-n_{l_3\downarrow}(\textbf{k}_2-\textbf{q})}{\omega + i\delta + \tilde{\epsilon}_{l_4,\uparrow}(\textbf{k}_2) - \tilde{\epsilon}_{l_3,\downarrow}(\textbf{k}_2-\textbf{q}) } \phi^{a*}_{l_4\uparrow}(\textbf{k}_2)\phi^{b}_{l_3\downarrow}(\textbf{k}_2-\textbf{q})\phi^{c}_{l_4\uparrow}(\textbf{k}_2)\phi^{d*}_{l_3\downarrow}(\textbf{k}_2-\textbf{q})\label{eq:chi0}
\end{equation}
\end{widetext}
is the susceptibility associated with the single-particle Hamiltonian $ H^{0,HF}$, which may be viewed as a $4\times4$ matrix written in the basis $AA,BB,AB,BA$.

Finally, we write Eq.~(\ref{eq:chi}) in matrix form and relate it to the physical response function in Eq.~(\ref{eq:physresp}), yielding
\begin{equation}
\chi_{\tau}(\textbf{q},\omega) = -\text{Tr}'\left[ \Big(1-2U_0\chi_0(\textbf{q},\omega)\Big)^{-1}\chi_0(\textbf{q},\omega)\right].\label{eq:finalchi}
\end{equation}
In this equation, all the matrices are $4\times4$. but the $\text{Tr}'$ is taken only over the ``diagonal'' elements, $\text{Tr}'\chi^{ab,cd}=\sum_{a,c=A,B}\chi^{aa,cc}$. Eq.~(\ref{eq:finalchi}) is one of our main results.

When $\text{Im}\chi(\textbf{q},\omega) \ne 0$ the system may absorb energy from a perturbation
that flips an electron spin,
so that the system has spin excitations with energy $\omega$ at momentum $q$; as a function
of $\omega$ for fixed $q$ this either comes over a range of frequencies, where there
is a continuum of excitations, or as sharp poles where there is a collective mode \cite{Giuliani_book}.
The latter case is characterized by $\text{Det}(1-2U_0\chi_0(\mathbf{q},\omega))=0$.
An example of $\chi(\mathbf{q},\omega)$ is illustrated in Fig. \ref{fig:phys_response},
where both a continuum and a sharp collective mode are evident.  Fig.~\ref{fig:chi}
shows the same example on a linear scale. In this case a sharp collective mode is expected at the point where the relevant determinant vanishes.
This mode is separated from the ``incoherent'' particle-hole excitations whose edge is denoted by $\omega_c$.

In addition to the collective mode that is evident in Fig. \ref{fig:chi}, a second mode arises
very close to the particle-hole continuum edge, which is rather difficult to discern in the
response function due to its close proximity to the continuum excitations. The presence of
such a mode can be demonstrated explicitly by examining the low hole-doping limit.
We now turn to this discussion.

\begin{figure}
	\includegraphics[width=0.45\textwidth]{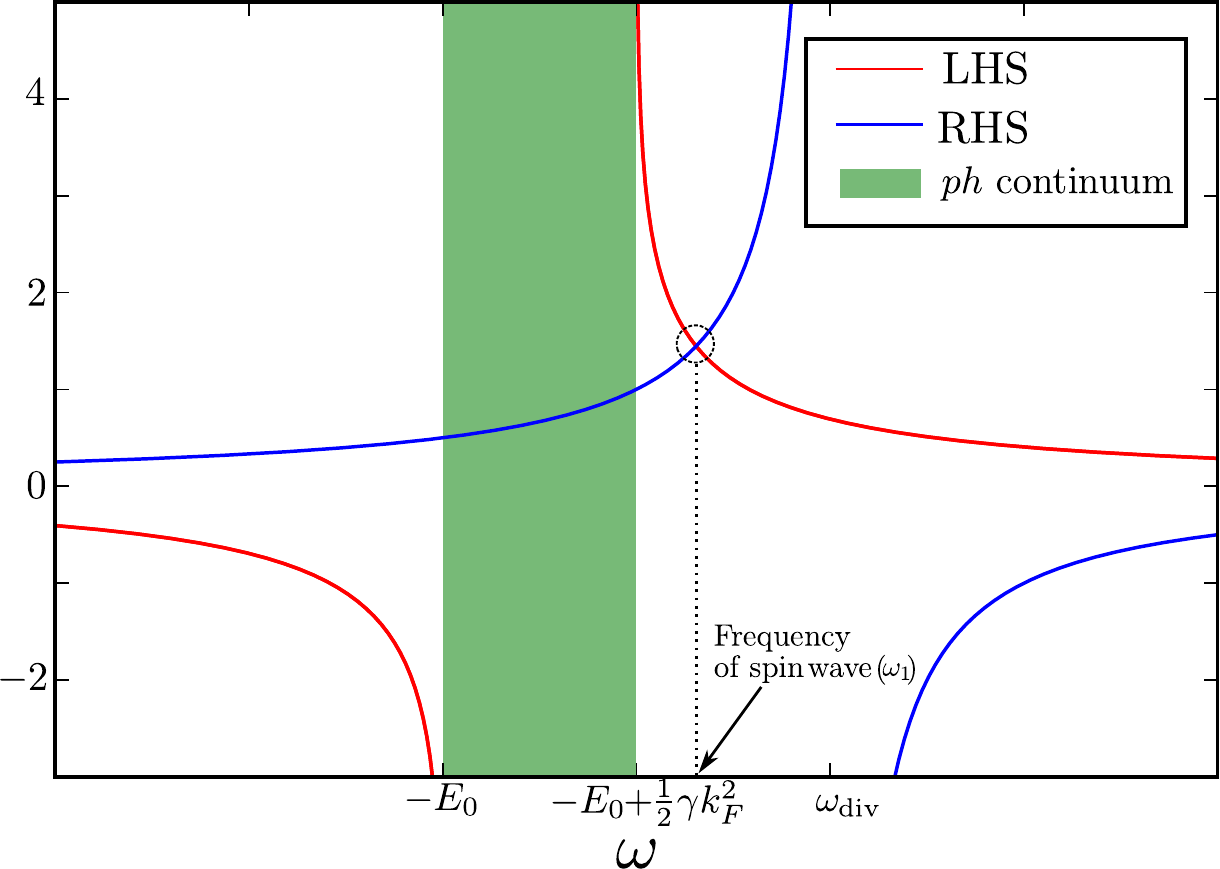}
	\caption{Schematic representation of the left and right hand sides of Eq.~(\ref{eq:SW12}) as functions of $\omega$, shown in red and blue respectively. For low enough $k_F$, an isolated spin wave mode is always present.}\label{fig:pole}
\end{figure}

\section{Spin-wave modes for small hole-doping}
For small densities of holes, it is possible to make analytical progress on finding zeros
of $\text{Det}(1-2U\chi_0(\mathbf{q},\omega))$ in the limit $q \rightarrow 0$,
indicating the location of sharp, collective spin-wave modes.
Specifying $\tau=1$ as the valley we will focus upon,
the valence bands are indexed by $\alpha=-1$ in Eq.~(\ref{eq:HFE}).
The dominant contributions to $\chi_0$ in Eq.~(\ref{eq:chi0})
come from $l_3 = l_4 = \{\tau=1,\alpha=-1\}$. This leads to the approximate expression
\begin{align}
 \tilde{\chi}_0^{ab,cd}(\textbf{q}=0) &=  - \frac{1}{L_xL_y}\sum_{\textbf{k}} M^{ab,cd}(\textbf{k}) \frac{\Delta n(k)}{\omega + i\delta + \Delta\tilde{\epsilon}(k)  },
\end{align}
where $\Delta n(k) = n_{\uparrow}(k) - n_{\downarrow}(k)$ and $\Delta\tilde{\epsilon}(k) = \lambda - (\tilde{m}_\uparrow-\tilde{m}_\downarrow) - U(n_\uparrow(k)-n_\downarrow(k)) - \frac{1}{2}\left(\frac{1}{\tilde{m}_\uparrow} - \frac{1}{\tilde{m}_\downarrow}\right)(atk)^2 \equiv E_0 - \frac{1}{2}\gamma k^2$, where $E_0 = \lambda -(\tilde{m}_{\uparrow} - \tilde{m}_{\downarrow}) - U_0 (n_{\uparrow}(k)-n_{\downarrow}(k))$ and $\gamma = \left(\frac{1}{\tilde{m}_\uparrow} - \frac{1}{\tilde{m}_\downarrow}\right)(at)^2$.
Notice we have employed a small $k$ expansion of $\tilde\epsilon(k)$, which works well because $\Delta n(k)$
differs from zero only at small $k$ in the low hole doping limit.
The particle-hole continuum is identified by the interval of $\omega$ for which $\omega + \Delta \tilde{\epsilon}(k)$
vanishes for some $k$ where $\Delta n(k) \ne 0$. This range is given in the present
approximation by $-E_0<\omega<-E_0 + \frac12\gamma k_F^2 \equiv \omega_c$,
where $k_F$ is the Fermi wavevector for the pocket of holes in the valence band.

The matrix elements $M^{ab,cd}(\mathbf{k}) =  \phi^{a*}_{\uparrow}(\textbf{k})\phi^{b}_{\downarrow}(\textbf{k})$ can be obtained by similarly expanding the Hartree-Fock wave functions for small $k$,
\begin{equation}
\tilde{\phi}_s(\textbf{k})
\approx \left[
\begin{array}{c}
e^{-i\phi} \frac{atk}{2\tilde{m}_s} \\
-[1 - \frac{(atk)^2}{8\tilde{m}_s^2}  ]\\
\end{array}\right],
\end{equation}
where only up to second order terms in $k$  are kept.
To this order the only relevant non-vanishing elements of the $M$ matrix are
\begin{align}
M^{AA,BB} &= M^{BB,AA}= \frac{(atk)^2}{4\tilde{m}_\uparrow\tilde{m}_\downarrow}, \nonumber \\
M^{BB,BB} &= 1 - \frac{(atk)^2}{4\tilde{m}_\uparrow^2}- \frac{(atk)^2}{4\tilde{m}_\downarrow^2}, \nonumber \\
M^{AB,BA} &= M^{BA,AB}=\frac{(atk)^2}{4\tilde{m}_\uparrow^2}.\nonumber
\end{align}
Except for
$M^{AA,AA}$ which vanishes to $\mathcal{O}(k^2)$, all the other entries of $M$ contain phases of the form $e^{-i\phi}$,
with $\phi$ the angle of {\bf k} with respect to the $k_x$-axis, which
vanishes upon integration over momentum.  Thus these do not contribute to $\tilde{\chi}_0$. At $\textbf{q}=0$, $\tilde{\chi}_0$ has a block-diagonal form and $\text{Det}(1-2U\chi_0(\mathbf{q},\omega))$ can be written as the product of two subdeterminants, $D_1$ and $D_2$, given by
\begin{align}
D_1 =& (1-2U_0\tilde{\chi}_0^{AA,AA})(1-2U_0\tilde{\chi}_0^{BB,BB}) \nonumber\\
&~~~~~~~~~~~~- 4U_0^2\tilde{\chi}_0^{AA,BB}\tilde{\chi}_0^{BB,AA}, \\
D_2 =& 1 - 4U_0^2\tilde{\chi}_0^{AB,BA}\tilde{\chi}_0^{BA,AB}.
\end{align}
If either of these vanishes at an $\omega$ outside the particle-hole continuum frequency interval, there
is a sharp collective mode at that frequency.  Note that particular response functions appearing in
$D_1$ and $D_2$ indicate that the former is associated with spin flips in which electrons remain
in the same orbital, while the latter arises due to electrons which both flip spin and change
orbital.

Using the integrals
\begin{align}
I_0 &= \frac{1}{L_xL_y}\sum_{|\textbf{k}| < k_F}\frac{1}{\omega+E_0-\frac{1}{2}\gamma k^2} \nonumber\\
&= \int_0^{k_F} \frac{kdk}{2\pi}\frac{1}{\omega+E_0-\frac{1}{2}\gamma k^2}\nonumber \\
&= -\frac{1}{2\pi \gamma}\text{ln}\left(\frac{\omega+E_0-\frac{1}{2}\gamma k_F^2}{\omega+E_0}\right)
\end{align}
and
\begin{align}
I_1 &= \frac{1}{L_xL_y}\sum_{|\textbf{k}| < k_F}\frac{k^2}{\omega+E_0-\frac{1}{2}\gamma k^2}\nonumber \\
&= \frac{1}{2\pi\gamma}\left[-\frac{\omega+E_0}{\gamma}\text{ln}\left(\frac{\omega+E_0-\frac{1}{2}\gamma k_F^2}{\omega+E_0}\right)-k_F^2\right],
\end{align}
the condition $D_1 = 0$ reduces to
\begin{equation}
1-2U_0\left(I_0 - \frac{(at)^2}{4}\left(\frac{1}{\tilde{m}_\uparrow^2} + \frac{1}{\tilde{m}_\downarrow^2}\right)I_1\right) = \frac{U_0^2 (at)^4}{4\tilde{m}_\uparrow^2\tilde{m}_\downarrow^2}.\label{eq:SW1}
\end{equation}
Similarly, $D_2 = 0$ can be simplified to
\begin{equation}
I_1 = \pm \frac{2\tilde{m}_\uparrow\tilde{m}_\downarrow}{U_0(at)^2}.\label{eq:SW2}
\end{equation}

\begin{center}
	\begin{figure*}[t]
		\includegraphics[width=0.95\textwidth]{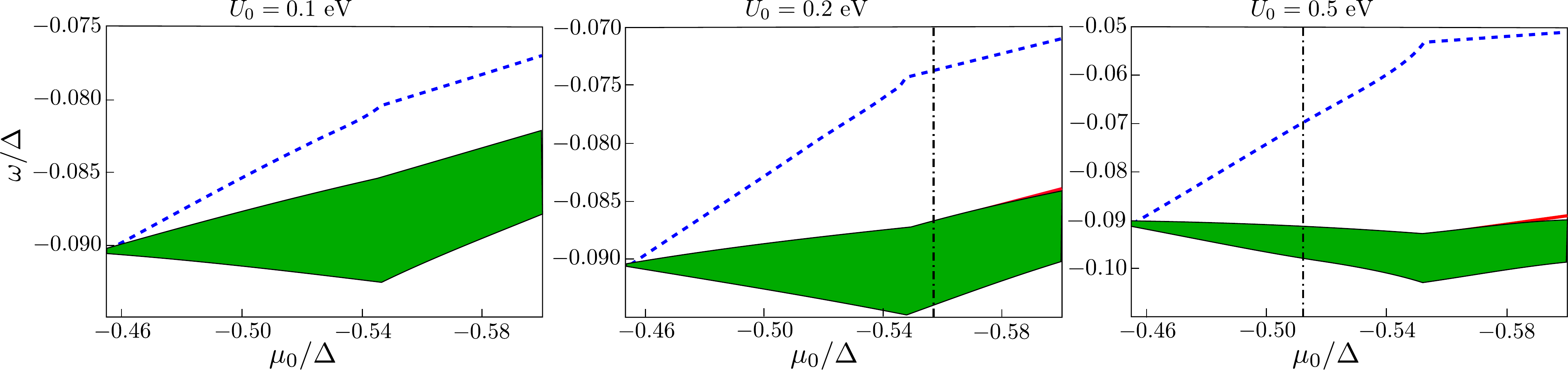}
		\caption{Spin wave excitations and the particle-hole continuum as a function of the chemical potential ($\mu_0$) shown for three different values of the interaction strength $U_0$ when $\mathbf{q}=\mathbf{0}$. The green band corresponds to the particle-hole continuum as is shown in Fig.~\ref{fig:chi}. The blue dashed line corresponds to the isolated mode at frequency $\omega_1$ described in Fig.~\ref{fig:chi} and Fig.~\ref{fig:pole}. The mode corresponding to Eq.~(\ref{eq:2ndmode}) is barely visible as a red line. The vertical lines indicate the boundary beyond which the stability condition is violated (see main text for details).}\label{fig:withmu}
	\end{figure*}
\end{center}

The condition Eq.~(\ref{eq:SW1}) will be met for some value of $\omega$ outside
the particle-hole continuum, for small interaction strength $U_0$.
This can be understood as follows.
For small $U_0$, we approximate the equation as
\begin{equation}
\frac{(at)^2}{4}\left(\frac{1}{\tilde{m}_\uparrow^2} + \frac{1}{\tilde{m}_\downarrow^2}\right)I_1 \approx I_0 -\frac{1}{2U_0}.
\end{equation}
Using the fact that
$$
I_1=\frac{\omega+E_0}{\gamma}I_0-\frac{k_F^2}{2\pi\gamma}
$$
this equation can be recast as
\begin{align}
I_0 = \frac{(at/2)^2\left(\frac{1}{\tilde{m}_\uparrow^2} + \frac{1}{\tilde{m}_\downarrow^2}\right)\frac{k_F^2}{2\pi\gamma}-\frac{1}{2U_0}}{(at/2)^2\left(\frac{1}{\tilde{m}_\uparrow^2} + \frac{1}{\tilde{m}_\downarrow^2}\right)\frac{E_0+\omega}{\gamma}-1}.\label{eq:SW12}
\end{align}
The numerator of the right hand side of this equation is negative for
small $U_0$.  As $\omega$ increases from large negative values, the right hand side is
positive and increases in magnitude, diverging at
\begin{align}
\omega=\omega_{\text{div}} \equiv -E_0 + \frac{4\gamma}{a^2t^2} \left(\frac{1}{\tilde{m}_\uparrow^2} + \frac{1}{\tilde{m}_\downarrow^2}\right)^{-1}.
\end{align}
Importantly, $\omega_{\text{div}} > \omega_c$ in the low doping limit, so the divergence
is above the particle-hole continuum.
Above $\omega_{div}$ the right hand side increases uniformly from arbitrarily large negative values,
eventually vanishing at large positive $\omega$.
By contrast, $I_0$ diverges to large negative values as $\omega \rightarrow -E_0$ from below, and
comes down from arbitrarily large positive values starting at the particle-hole continuum edge
$\omega_c$.  This guarantees there will be a crossing of the left and right
hand sides of Eq.~(\ref{eq:SW12}) between this edge and $\omega_{\text{div}}$, and a collective mode with frequency $\omega_1$ in this interval.  This is qualitatively shown in Fig.~\ref{fig:pole}.
Note that for decreasing $U_0$ this solution moves closer to the particle-hole continuum, which we indeed find numerically, as illustrated in Fig.~\ref{fig:withmu}. As is shown in Appendix B, for small $U_0$ and small hole doping, one can show that for $\textbf{q}=0$
\begin{align}
  \omega_1 \approx -E_0+ \frac12\gamma k_F^2\left(1+e^{-\pi\gamma/U_0}\right) \label{eq:1stmode}.
\end{align}

The second condition Eq.~(\ref{eq:SW2}), for small $U_0$, can only be satisfied for the negative sign of the right hand side. The position of the spinwave mode at $\mathbf{q}=0$ can be approximately evaluated to be
\begin{align}
\omega_2 \approx -E_0+ \frac12\gamma k_F^2\left(1+e^{-\epsilon_0/k_F^2U_0}\right), \label{eq:2ndmode}
\end{align}
where $\epsilon_0 = 8\pi \tilde{m}_{\uparrow}\tilde{m}_{\downarrow}/a^2t^2$. 
It is clear from Eqs.~(\ref{eq:1stmode}) and (\ref{eq:2ndmode}) that the separation of $\omega_2$ from the particle-hole continuum is very small when compared to that of $\omega_1$ for small hole doping and for the relevant parameter range. This result is again consistent with our numerical solutions, as illustrated in Fig.~\ref{fig:withmu}.

We conclude this section with two comments on these results. First, the appearance
of a sharp collective mode with arbitrarily small $U_0$ supports the interpretation
of the non-interacting groundstate as being effectively polarized in a ``pseudospin''
spin variable, $\sigma_z\tau_z$, as discussed in the Introduction.  When interactions
are introduced, incoherent particle-hole excitations are pushed up in energy
via a loss of exchange energy which, for repulsive interactions, generically lowers
the groundstate energy for a polarized state.  However, an appropriate linear combination
of particle-hole pair states can minimize this loss of exchange energy, leading to
the sharp collective mode.

Secondly, although we have demonstrated the existence of two discrete modes, the second of these
(at $\omega=\omega_2$) lies exceedingly close to the particle-hole continuum edge.  This means that
small perturbations can easily admix these different kinds of modes together, making the
detection of the second mode challenging.  Indeed, in our own numerics the introduction of
broadening in our discrete wavevector sum, introduced to simulate the thermodynamic limit,
typically mixes this mode with the continuum.  In this situation the mode does not show
up sharply in the response function we focus upon.  We note that our analysis shows the
mode to be associated with simultaneous spin flip {\it and} a change of orbital, $A \leftrightarrow B$,
so that we expect this second mode should show up more prominently in more complicated
response functions that simultaneously probe both of these.


\section{Numerical Results and Discussion}
In general,
to compute $\chi_\tau$ we need to know $\chi_0$.  This can be obtained numerically, and we accomplish this by approximating the integral in Eq.~(\ref{eq:chi0}) as a discrete sum.
For our calculations we discretize
momenta onto a $100\times 100$ two dimensional grid,
with each momentum component running from $-k_0$ to $+k_0$. We have checked that the contribution to
$\chi_0$ dies off quickly within the range of momentum integration. We also discretize $\omega$ to a set
of 5000 points, within which we compute physical response functions. A small but non-vanishing imaginary
$\eta$ is retained, of the order of the spacing of the $\omega$ values, to produce the continuity
expected in the thermodynamic limit (where the momentum grid over which we sum becomes
arbitrarily fine). Figs. \ref{fig:phys_response} and \ref{fig:chi} depict typical results.

%
\begin{figure}
	\includegraphics[width=0.45\textwidth]{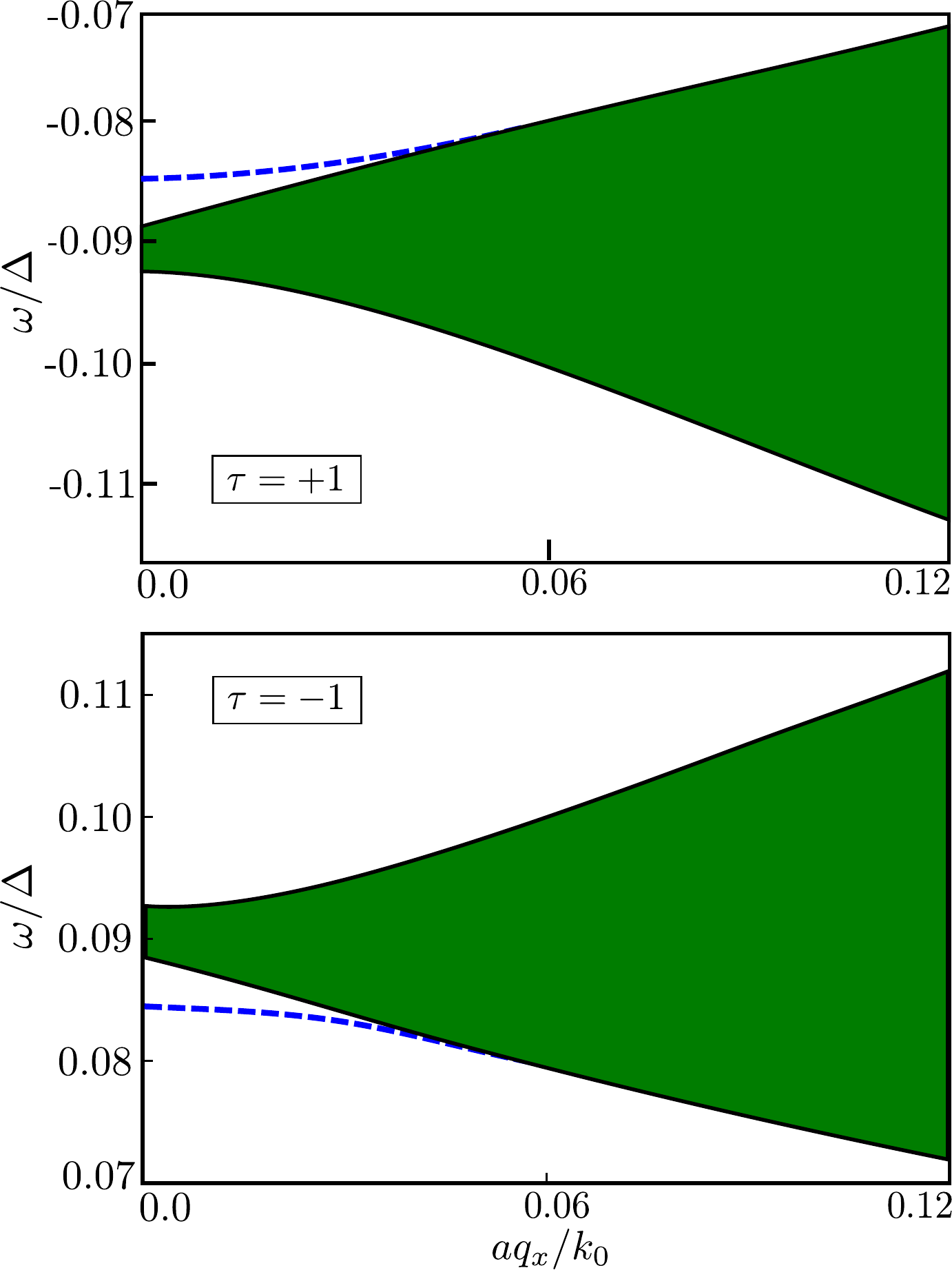}
	\caption{The blue line depicts the dispersion of the isolated spin wave excitation,ie, the $\omega,q_x$ points for which the real part of the denominator of the spin susceptibility given by Eq.~(\ref{eq:finalchi}) vanishes. The green continuum represents the particle-hole excitations for which the denominator of Eq.~(\ref{eq:finalchi}) has a nonvanishing imaginary component as is shown in Fig.~\ref{fig:chi}. Here we have taken $U_0 = 0.2$eV.}\label{fig:U1}
\end{figure}

The response function Eq.~(\ref{eq:corr}) qualitatively describes the dynamics of an electron-hole pair
between bands of opposite spins. The lowest energy excitations necessarily
involve the bands nearest the chemical potential $\mu$.
When $\mu$ is within the gap so that the system is insulating,
such an excitation will have energy comparable to the band gap
$E_g \sim 1$eV \cite{Ross_2013,Ugeda_2014,Wu_2015}.
On the other hand,
when hole-doped, the
chemical potential falls below the top of the valence band, electron-hole pairs
from the two spin species in the valence band become available (see Fig.~\ref{fig:band}).
The resulting excitations can have energy of order $\lambda \sim 0.1$eV, a considerably
lower energy scale.
Discrete poles
in $\chi$ have
infinite lifetime and represent the collective spin-wave modes of the system; these only
can arise when interactions are included in the model.   A set of representative plots
illustrating both the
spin-wave dispersion and the particle-hole continuum
are shown in Fig.~\ref{fig:U1} for both the valleys.
Note the clear symmetry apparent between the two valley responses when $\omega \rightarrow -\omega$.
This is a manifestation of time-reversal symmetry, and indicates that strong absorption
from a perturbation with one helicity in one of the two valleys implies equally strong
absorption in the other valley when the helicity is reversed.

It is interesting to consider the possible consequences of this if the system develops true
ferromagnetism, which is thought to occur above some critical interaction strength $U_c$
\cite{Scrace_2015,Braz_2017}.  In the simplest description, this leads to different
self-consistent exchange fields and different hole populations for each valley \cite{Braz_2017}.
The computation of spin-response in this situation
is essentially the same as carried out in our study, but the effective chemical potential would
be different for each valley.  In this case we expect the spin response to be different for
the two possible perturbations, reflecting the broken time-reversal symmetry in the groundstate.
Such behavior has indeed been observed for {\it electron}-doped
TMD's \cite{Scrace_2015}.

Another feature apparent in Fig.~\ref{fig:withmu} is
a cusp in the continuum spectrum, which appears at
$\mu_0=\mu_c\approx -0.55\Delta$. This is the point at which the chemical potential touches the top of the
lower valence band (Fig.~\ref{fig:band}). For $\mu_0>\mu_c$,
a particle-hole
continuum is only present at non-vanishing
frequencies determined by the difference in energy between the highest occupied and the lowest unoccupied bands of opposite spins. However, for $\mu_0 \leq \mu_c$, low energy particle-hole
excitations set in for processes in which (for one of the valleys) a spin-down valence band
electron is excited to the spin-up valence band at finite wave vector, but vanishing frequencies. This is further illustrated in Fig.~\ref{fig:loww}, in which one finds the continuum excitations
reaching down to zero energy, at a finite $q_x$, only when the chemical potential is below this critical value.

As is apparent from Fig.~\ref{fig:chi} the first spin-wave mode from the condition Eq.~(\ref{eq:SW12}) appears above the continuum. Further, for a given $U_0$, the separation from the continuum increases linearly with increasing hole doping, as illustrated in Fig.~\ref{fig:U1},
until the chemical potential touches the top of the lower valence band. At this point a similar cusp as for the continuum appears in the spin wave dispersion. The linear increase of the separation between the spin wave mode
and the top of the particle-hole continuum at small hole doping can be understood in the following way.
As shown in Appendix B, Eq.~(\ref{eq:SW12}) can be approximated for small hole doping and small $U_0$ by
\begin{align}
\frac{\delta_0}{\delta_0 + c_0\delta \mu} \approx e^{-\pi\gamma/U}, \label{eq:2ndmodeexpansion}
\end{align}
where $\delta_0$ is the separation of the spin wave from the continuum, and $\delta\mu$
is the change in chemical potential due to hole doping, and the constant $c_0 = \gamma/\tilde{m}_\uparrow$.
As the right hand side of the equation is independent of $\delta\mu$, the solution
$\delta_0$ should also be proportional to $\delta\mu$.

\begin{center}
	\begin{figure}[t]
		\includegraphics[width=0.45\textwidth]{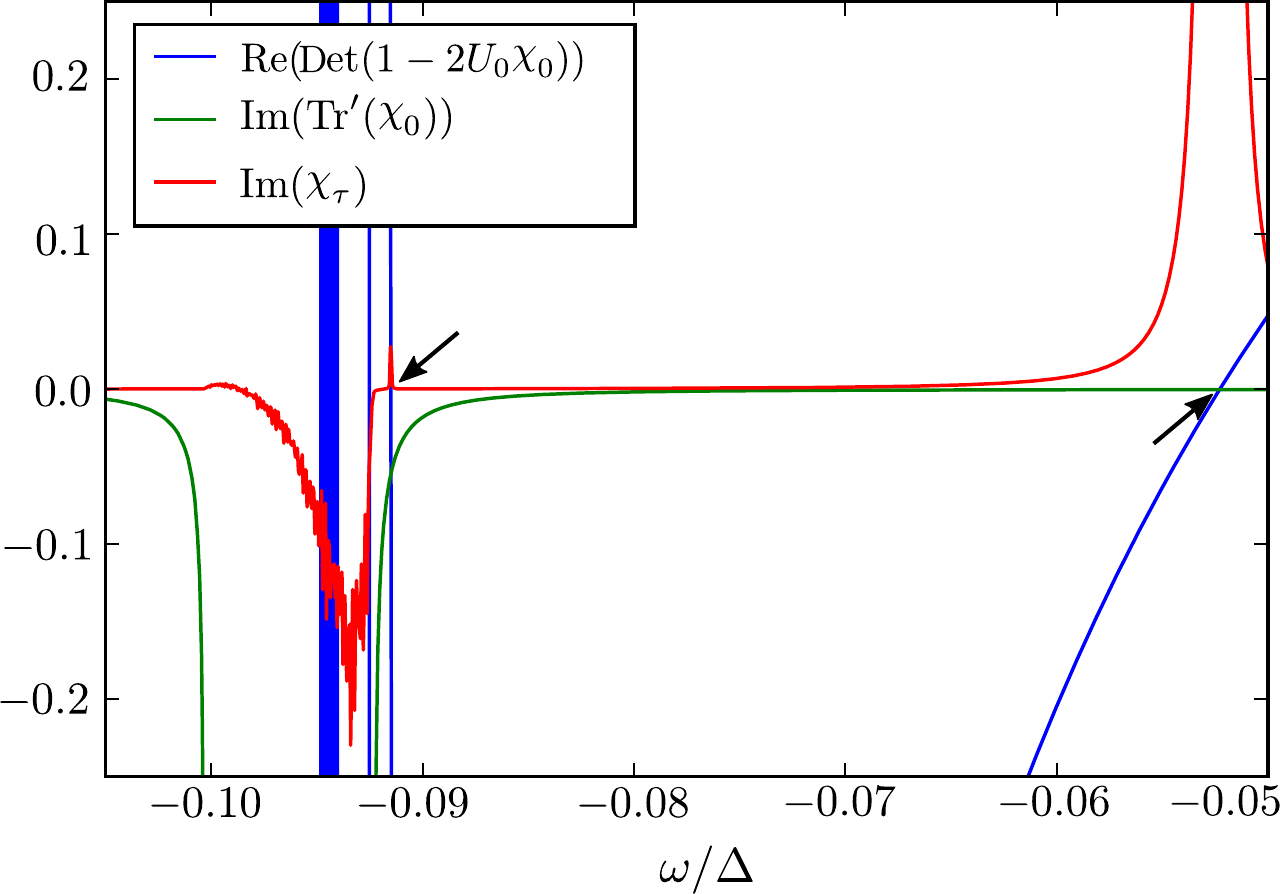}
		\caption{Spin susceptibility for $U_0 = 0.5$eV, $\tau=+1$ and $\mu_0 = -0.57\Delta$. Two discrete spin wave modes (indicated by arrows) are visible near $\omega = -0.055\Delta$ and $\omega=-0.092\Delta$, with the second mode very close to the continuum. However, the positivity $\omega (-\text{Im}\chi)>0$ does not hold for all $\omega$ implying that our assumed Hartree-Fock state is not the true ground state.}\label{fig:2ndsw}
	\end{figure}
\end{center}

As discussed in the previous section, the second spin wave solution of Eq.~(\ref{eq:2ndmode}) lies extremely close to the continuum, and so is almost invisible in our numerical solutions for the range of the parameters we
consider. One expects this mode to be visible for larger $U_0$ and larger hole doping. However, in our calculations
we find that the stability condition\cite{Giuliani_book} $\omega (-\text{Im}\chi_\tau)>0$ fails for some range of $\omega$
for $U_0$ large enough that we
are able to numerically resolve the mode from the continuum.  An example of this is shown in Fig.~\ref{fig:2ndsw}. The point beyond which this stability condition is not satisfied is indicated by vertical lines in Fig.~\ref{fig:withmu}.
Note that, physically,
the instability we find in the response functions indicates that the symmetry of the
ground-state we are assuming is broken, very likely into a state with inter-orbital coherence.
Whether such a state exists at large $U$, or is preempted by a first-order transition into
a state with different hole populations in the valleys, requires a more general Hartree-Fock
study than we have presented in this work, and is left for future study.

\section{Summary}

In this paper, we have studied collective excitations of a simple TMD model, showing that even
without the formation of spontaneous magnetic order, interactions induce sharp collective
modes that are commonly associated with such order.  The presence of these modes can be
understood as a consequence of intrinsic order induced by the strong spin-orbit interaction
that yields different energetic orderings of spins in different valleys, and arises when
the system is doped.  The presence of these modes is a direct analog of ``Silin-Leggett modes''
present in a simple Fermi liquid subject to a magnetic field, such that the Fermi wavevector
becomes spin-dependent.  Our analysis is developed using the time-dependent Hartree-Fock
approximation of a physical spin response function,
and reveals two sharp modes in addition to a continuum of particle-hole excitations.
While one of these modes (associated with spin flips for electrons maintaining their orbital
index)  breaks out from the continuum in a clear way, the other (associated with electrons
changing both spin and orbital) remains very close to the continuum edge and is difficult
to distinguish independently.  Signatures of how the subbands are populated can be seen in
properties of the spin response functions when the chemical potential is modified, which
in principle can be accomplished by gating the system.

Our calculations indicate that with strong enough interaction the system becomes unstable. Within our model this would likely be to a state with inter-orbital coherence, but first order instabilities in which the system spontaneously forms unequal valley and spin populations are also possible, which may preempt any instability indicated in linear response.  The validity of the simple model that we use, Eq.~(\ref{eq:ham}), is also limited by the positions of other bands in the system, notably, at the $\Gamma$ point~\cite{Xiao_2012}. For MoS$_2$, this separation is small as bands near the $\Gamma$ point lie 0.1-0.2 eV below the tops of the bands at the $K,K'$ points.  The separation in energy is larger for certain dichalcogenides, such as WS$_2$, MoSe$_2$, WSe$_2$, MoTe$_2$, WTe$_2$, among others.  Our results, which are based on a simple two-band model near $K,K'$ points, will change qualitatively when the Fermi energy is low enough that bands at the $\Gamma$ points contain holes. Whatever the true groundstate of the system, our formalism in principle allows a calculation of the density matrix associated with it, and of collective modes around it.  Moreover, the approach we present can be extended to more general response functions (for example, involving spin and orbital simultaneously) which could reveal further and perhaps clearer signatures of the two collective modes we find in our analysis.  Exploration of these represent interesting directions for future work.

{\it Acknowledgements} -- HAF acknowledges
the support of US-National Science Foundation through grant nos. DMR-1506263 and DMR-1506460,
and the US-Israel Binational Science Foundation through grant no. 2016130.  HAF also
thanks Aspen Center for Physics (NSF grant PHY-1607611), where part of this
work was performed. The research of DKM was supported in part by the INFOSYS scholarship for senior students. A. K. acknowledges the support from the Indian Institute of Technology - Kanpur.

\appendix

\section{Details of Time dependent Hartree-Fock Approximation}
In this Appendix we provides a few details of the calculation leading to Eq.~(\ref{deriv}). The equation of motion of $\tilde{\chi}$, Eq.~(\ref{eq:chitilde}), is
\begin{widetext}
	\begin{align}
    i\partial_t\tilde{\chi}_{l_1l_2l_3l_4}(\textbf{k}_1\textbf{k}_2\textbf{q},t) = & \{n_{l_1\uparrow}(\textbf{k}_1+\textbf{q})-n_{l_2\downarrow}(\textbf{k}_1)\}\delta_{l_1l_4}\delta_{l_2l_3}\delta_{\textbf{k}_1,\textbf{k}_2-\textbf{q}}  +  i\Theta(t)\langle\Big[[H_0,c^\dagger_{l_1\uparrow}(\textbf{k}_1+\textbf{q})c_{l_2\downarrow}(\textbf{k}_1)](t),c^\dagger_{l_3\downarrow}(\textbf{k}_2-\textbf{q})c_{l_4\uparrow}(\textbf{k}_2)\Big]\rangle \nonumber \\
                                                                                   &+ i\Theta(t)\langle\Big[[H_{\text{int}},c^\dagger_{l_1\uparrow}(\textbf{k}_1+\textbf{q})c_{l_2\downarrow}(\textbf{k}_1)](t),c^\dagger_{l_3\downarrow}(\textbf{k}_2-\textbf{q})c_{l_4\uparrow}(\textbf{k}_2)\Big]\rangle .
                                                                                     \label{eomchi}
  \end{align}
The first commutator reads
  \begin{align}
    {[}H_0,c^\dagger_{l_1\uparrow}(\textbf{k}_1+\textbf{q})c_{l_2\downarrow}(\textbf{k}_1)] =& \sum_lh^0_{ll_1,\uparrow}(\textbf{k}_1+\textbf{q})c^\dagger_{l\uparrow}(\textbf{k}_1+\textbf{q})c_{l_2\downarrow}(\textbf{k}_1) - \sum_{l'}h^0_{l_2l',\downarrow}(\textbf{k}_1)c^\dagger_{l_1\uparrow}(\textbf{k}_1+\textbf{q})c_{l'\downarrow}(\textbf{k}_1).
                                                                                               \label{H0comm}
  \end{align}
The first commutator appearing in the last term of Eq.~(\ref{eomchi}) is
  \begin{align}
      \Big[H_{\text{int}},& c^\dagger_{l_1\uparrow}(\textbf{k}_1+\textbf{q})c_{l_2\downarrow}(\textbf{k}_1)\Big] \nonumber \\
    = & 2 U \sum_{\{l_i,\textbf{k}_i\}} \Big[ f_{l_6l_7,\uparrow\uparrow}(\textbf{k}_6,\textbf{k}_6+\textbf{q}')f_{l_5l_8,\uparrow\uparrow}(\textbf{k}_5,\textbf{k}_5-\textbf{q}')   c^\dagger_{l_5\uparrow}(\textbf{k}_5)c^\dagger_{l_6\uparrow}(\textbf{k}_6)c_{l_7\uparrow}(\textbf{k}_6+\textbf{q}')c_{l_2\downarrow}(\textbf{k}_1) \delta_{l_1,l_8}\delta_{\textbf{k}_5-\textbf{q}',\textbf{k}_1+\textbf{q}} \nonumber \\
    & + f_{l_6l_7,\downarrow\downarrow}(\textbf{k}_6,\textbf{k}_6+\textbf{q}')f_{l_5l_8,\downarrow\downarrow}(\textbf{k}_5,\textbf{k}_5-\textbf{q}')   c^\dagger_{l_1\uparrow}(\textbf{k}_1+\textbf{q})c^\dagger_{l_5\downarrow}(\textbf{k}_5)c_{l_7\downarrow}(\textbf{k}_6+\textbf{q}')c_{l_8\downarrow}(\textbf{k}_5-\textbf{q}') \delta_{l_2,l_6}\delta_{\textbf{k}_1,\textbf{k}_6} \nonumber \\
     & - f_{l_6l_7,\uparrow\uparrow}(\textbf{k}_6,\textbf{k}_6+\textbf{q}')f_{l_5l_8,\downarrow\downarrow}(\textbf{k}_5,\textbf{k}_5-\textbf{q}')  c^\dagger_{l_5\downarrow}(\textbf{k}_5)c^\dagger_{l_6\uparrow}(\textbf{k}_6)c_{l_8\downarrow}(\textbf{k}_5-\textbf{q}')c_{l_2\downarrow}(\textbf{k}_1) \delta_{l_1,l_7}\delta_{\textbf{k}_1+\textbf{q},\textbf{k}_6+\textbf{q}'} \nonumber \\
    & - f_{l_6l_7,\uparrow\uparrow}(\textbf{k}_6,\textbf{k}_6+\textbf{q}')f_{l_5l_8,\downarrow\downarrow}(\textbf{k}_5,\textbf{k}_5-\textbf{q}')  c^\dagger_{l_1\uparrow}(\textbf{k}_1+\textbf{q})c^\dagger_{l_6\uparrow}(\textbf{k}_6)c_{l_7\uparrow}(\textbf{k}_6+\textbf{q}')c_{l_8\downarrow}(\textbf{k}_5-\textbf{q}')\delta_{l_2,l_5}\delta_{\textbf{k}_1,\textbf{k}_6} \Big].
  \end{align}
Here, for notational simplicity, we have absorbed the $\delta_{\tau_i\tau_j}$ factors inside the $f_{l_il_j}$s. We next employ the Hartree-Fock approximation and find that the $\textbf{q}'=0$ terms cancel each other. The other terms are
  \begin{align}
    \Big[H_{\text{int}}, c^\dagger_{l_1\uparrow}(\textbf{k}_1+\textbf{q})&c_{l_2\downarrow}(\textbf{k}_1)\Big] \nonumber \\
    \rightarrow -2U  \sum_{\{l_i,\textbf{k}_i\}}\Big[ & f_{l_6l_5,\uparrow\uparrow}(\textbf{k}_1+\textbf{q},\textbf{k}_1+\textbf{q}+\textbf{q}')f_{l_5l_1,\uparrow\uparrow}(\textbf{k}_1+\textbf{q}+\textbf{q}',\textbf{k}_1+\textbf{q}) n_{l_5\uparrow}(\textbf{k}_1+\textbf{q}+\textbf{q}')c^\dagger_{l_6,\uparrow}(\textbf{k}_1+\textbf{q})c_{l_2,\downarrow}(\textbf{k}_1) \nonumber \\
    + & f_{l_2l_5,\downarrow\downarrow}(\textbf{k}_1,\textbf{k}_1+\textbf{q}')f_{l_5l_8,\downarrow\downarrow}(\textbf{k}_1+\textbf{q}',\textbf{k}_1)  n_{l_5\downarrow}(\textbf{k}_1+\textbf{q}')c^\dagger_{l_1,\uparrow}(\textbf{k}_1+\textbf{q})c_{l_8,\downarrow}(\textbf{k}_1) \nonumber  \\
    - & f_{l_6l_1,\uparrow\uparrow}(\textbf{k}_1+\textbf{q}-\textbf{q}',\textbf{k}_1+\textbf{q})f_{l_2l_8,\downarrow\downarrow}(\textbf{k}_1,\textbf{k}_1-\textbf{q}')  n_{l_2\downarrow}(\textbf{k}_1)c^\dagger_{l_6,\uparrow}(\textbf{k}_1+\textbf{q}-\textbf{q}')c_{l_8,\downarrow}(\textbf{k}_1-\textbf{q}') \nonumber\\
    + & f_{l_2l_8,\downarrow\downarrow}(\textbf{k}_1,\textbf{k}_1-\textbf{q}')f_{l_6l_1,\uparrow\uparrow}(\textbf{k}_1+\textbf{q}-\textbf{q}',\textbf{k}_1+\textbf{q})  n_{l_1\uparrow}(\textbf{k}_1+\textbf{q})c^\dagger_{l_6,\uparrow}(\textbf{k}_1+\textbf{q}-\textbf{q}')c_{l_8,\downarrow}(\textbf{k}_1-\textbf{q}')  \Big] .
  \end{align}
At this point, we would like to point out that because $f_{l_il_j}\propto \delta_{\tau_i\tau_j}$ and $\tau_1=\tau_2$, all the electronic operators have the same valley index $\tau$ in this expression.

  Finally, we introduce $ \rho^{ab}_{s_1s_2}(\textbf{q}) = \sum_{ll'\textbf{k}}\phi^{a*}_{ls_1}(\textbf{k}+\textbf{q})c^\dagger_{l\uparrow}(\textbf{k}+\textbf{q})c_{l'\downarrow}(\textbf{k})\phi^b_{ls_2}(\textbf{k})$ and $n_{s}^{ab} = \sum_{\textbf{k}l}\phi^a_{ls}(\textbf{k})n_{ls}(\textbf{k})\phi^{b*}_{ls}(\textbf{k})$ to write
  \begin{align}
    {[}H_{\text{int}},&c^\dagger_{l_1\uparrow}(\textbf{k}_1+\textbf{q})c_{l_2\downarrow}(\textbf{k}_1)]\nonumber \\ \rightarrow &-2U\sum_{abl'}\Big[ n^{ab}_\uparrow\phi^b_{l_1\uparrow}(\textbf{k}_1+\textbf{q})\phi^{a*}_{l'\uparrow}(\textbf{k}_1+\textbf{q})c^\dagger_{l'\uparrow}(\textbf{k}_1+\textbf{q})c_{l_2\downarrow}(\textbf{k}_1) - n^{ab}_\downarrow\phi^b_{l'\downarrow}(\textbf{k}_1)\phi^{a*}_{l_2\downarrow}(\textbf{k}_1)c^\dagger_{l_1\uparrow}(\textbf{k}_1+\textbf{q})c_{l'\downarrow}(\textbf{k}_1) \Big] \nonumber \\
                             & + 2U\sum_{ab}\phi^a_{l_1\uparrow}(\textbf{k}_1+\textbf{q})\Big[n_{l_1\uparrow}(\textbf{k}_1+\textbf{q}) - n_{l_2\downarrow}(\textbf{k}_1) \Big] \phi^{b*}_{l_2\downarrow}(\textbf{k}_1)\rho^{ab}_{\uparrow\downarrow}(\textbf{q}).
                               \label{HIcomm}
  \end{align}
\end{widetext}
Substituting Eq.~(\ref{H0comm}) and Eq.~(\ref{HIcomm}) in Eq.~(\ref{eomchi}) we obtain Eq.~(\ref{deriv}) of the main text.

\section{Small hole-doping}
In this Appendix, we supply some details underlying Eqs.~(\ref{eq:1stmode}) and (\ref{eq:2ndmodeexpansion}).
For small $U_0$, assuming that the renormalized masses $\tilde{m}_s$ to be close to their non-interacting
values, we can write
\begin{align}
  \frac{1}{\tilde{m}^2_\uparrow}+\frac{1}{\tilde{m}^2_\downarrow} &\approx (\frac{2}{\Delta-\lambda})^2 + (\frac{2}{\Delta+\lambda})^2  \approx \frac{8}{\Delta^2}.
\end{align}
Furthermore, we note
\begin{align}
  \frac{\gamma}{(at)^2} = \frac{1}{\tilde{m}_\uparrow} - \frac{1}{\tilde{m}_\downarrow} \approx 4\frac{\lambda}{\Delta^2}.
\end{align}
These allow Eq.~(\ref{eq:SW12}) for small $U_0$  and small hole doping to be written as:
\begin{align}
\frac{-1}{2\pi\gamma}\text{ln}\left(\frac{\omega - \omega_c}{\omega - \omega_c + \frac12\gamma k_F^2}\right) \approx \frac{\frac{k_F^2}{2\pi}-\frac{\lambda}{U_0}}{\omega - \omega_c -2\lambda +\frac12\gamma k_F^2},
\end{align}
where $\omega_c = -E_0+\frac12\gamma k_F^2$ is the boundary of the continuum of particle-hole
excitations. Moreover, again for small $U_0$, assuming the upper valence band to have spin up(which is the case for $\tau=+1$),
we can write the chemical potential as
$\mu_0 \approx -\frac12\Delta + \lambda - \frac12 \frac{(at)^2k_F^2}{m_\uparrow}$, so that the
change in $\mu_0$ due to hole doping can be written as
$\delta \mu = \frac12 \frac{(at)^2k_F^2}{m_\uparrow}$. Using this in the above equation we get
\begin{align}
\frac{-1}{2\pi\gamma}\text{ln}\left(\frac{\omega - \omega_c}{\omega - \omega_c +c_0\delta \mu}\right) \approx \frac{1}{2U_0},
\end{align}
where, for small $\delta\mu$, $\omega - \omega_c$ and $\frac12\gamma k_F^2$ are neglected compared to $\lambda$. As the right-hand side is independent of $\delta \mu$, the solution $\omega -\omega_c$ should also scale as $\delta \mu$.

When $U_0$ is small, the above equation can be solved for $\omega = \omega_1 \approx -E_0+ {1\over 2}\gamma k_F^2(1+e^{-\pi\gamma/U_0})$. Note that this result differs from that of Eq.~(\ref{eq:2ndmode}) in that $k_F^2$ appears in the exponential in the latter.  This renders $|\omega_2-\omega_c|$ much smaller than
$|\omega_1-\omega_c|$ in the low hole-doping limit.

\bibliography{TMDRefs}

\end{document}